\documentclass[letterpaper, 10 pt, conference]{ieeeconf}

\IEEEoverridecommandlockouts                              
\usepackage{cite}
\usepackage{graphicx}

\title{\LARGE \bf Cross Comparison of Synonym Graphs in A~Multi~Linguistic~Context}

\author{\authorblockN{Dorina Strori}
\authorblockA{Computer Engineering\\
Bogazici University\\
Email: dorina\_strori@yahoo.com}
\and
\authorblockN{Ahmet Bombaci}
\authorblockA{Computer Engineering\\
Bogazici University\\
Email: ahmetbombaci@yahoo.com}
\and
\authorblockN{Haluk Bingol}
\authorblockA{Computer Engineering\\
Bogazici University\\
Email: bingol@boun.edu.tr}}

\begin{document}

\maketitle
\thispagestyle{empty}
\pagestyle{empty}
%%%%%%%%%%%%%%%%%%%%%%%%%%%%%%%%%%%%%%%%%%%%%%%%%%%%%%%%%%%%%%%%%%%%%%%%%%%%%%%%%

\begin{abstract}
Language is one of the most important aspects of human cognition; it represents the way we think, act and communicate with each other. Each language has its own history, background, and form. A language represents a lot of important cultural aspects of the nation speaking it. Languages differ and so do cultures. In this paper we analyze cultural differences between East and West in a multi-linguistic context from a complex networks point of view.  There has been considerable work on the topic of cultural differences by psychologist and sociologist. Also studies on complex networks that make use of WordNet have been done, but until now there is no previous work that uses WordNets from different Eastern and Western languages as complex lexical networks in order to obtain possible differences or similarities between the cultures using those respective languages. Our work aims to do this. 
\end{abstract}

\section{Introduction}
Synonyms are a well-known and important part of a language. According to a definition by Leibniz, two concepts are synonymous if the substitution of one for another never changes the truth value of the sentence in which the substitution occurs.  But such a pure synonymy either doesn't exist at all, or is very rare. Instead a weakened version of this definition which considers synonyms as relative to the context in which they are used, applies in general, i.e. two concepts are synonyms in a context if the substitution of one for another does not change the truth value of the context \cite{Wordnet:Introduction}. 

People may use several words to express a certain concept, i.e. a concept can be represented by these several words known as synonyms. If the set of synonyms of a certain concept is large, we may well say that this concept is important to the people using it. When a synonym for a concept is created, it is not done simply for fun and is not cost free. Generally, people need to create it and when doing this, some cognitive processing is done by the brain. If people need to extend the synonym set of a concept, then this means that this specific concept is widely used and important to them.  Since each nation has its own language, culture and way of living, the needs for synonyms of a certain concept may differ from one culture to the other according to the importance given to concepts by those cultures.  

Thinking in this way, it could be possible to capture culture differences as well as similarities by examining synonym networks from different languages. Since East and West are considered to be two extremes of culture, it would be interesting and worth studying on different Eastern and Western languages. This was our motivation and aim at the start of our work. 

There has been some previous work on the topic of linguistic networks. For example, Ferrer \& Sole made use of the British National Corpus in constructing huge word networks in which the words are linked to each other if they are directly neighbors in a sentence \cite{Ferrer_Sole}. They found that the network were scale free and showed small world properties. Motter constructed synonym networks by using an English thesaurus and found small-world characteristics \cite{Motter}. Sigman \& Cecchi used WordNet to analyze English nouns \cite{Sigman_Cecchi}. They analyzed the network by using the relationships between concepts, such as hyponymy, meronomy, holonymy, antonymy etc. They found that the presence of polysemy dramatically changes the compactness of the network. Steyvers \& Tenenbaum made a study on three types of semantic networks: word associations, WordNet, and Roget's Thesaurus \cite{Steyvers_Tenenbaum}. They showed that these networks have a small-world structure and that the distributions of the number of connections follow power laws.

As far as we are concerned, no prior work has been done in analyzing the synonym graphs of different languages in attempt to find possible culture differences as well as similarities. In this aspect ours is a novice work. 

Since the beginning of the work, we aimed at studying on four different languages, two eastern and two western ones, but we had the possibility to work only on English and Turkish, each of them being a representative of West and East respectively. Actually we are planning to extend the work to other languages such as: Italian, Hindu, Arabic and Hebrew, whose electronic resources  we have recently obtained. One of the most important issues was the choice of the electronic resource. We decided to make use of WordNet facility whose English and Turkish versions were available to us. 

The explanation of WordNet and the reason why we chose it will be given in the following section.  The rest of the paper is organized  as follows : Section 2 consists of a brief introduction to WordNet, in section 3 we explain how we constructed the synonym graphs, section 4 consists of  the graph analysis results, in section 5 we give the drawn conclusions and at the end, the references used.
\section{Wordnet}
WordNet is an on-line lexical reference system whose design is based on psycholinguistic theories of human lexical memory. English nouns, verbs, and adjectives are organized into synonym sets each representing one underlying lexical concept. Synonym sets are linked by different relations, such as antonymy, meronomy, hyponymy \cite{Wordnet:Introduction,Fellbaum}. Its development began in 1985 at Princeton University under the direction of George A. Miller. Several versions of it have been made available through the years. The most recent version for Windows is 2.1, but we used the version 2.0 for Windows, because that same version of the Turkish WordNet was available to us at the time. The English 2.0 version contains 125207 words and 99143 synonym sets. On the other hand, Turkish WordNet was developped at Sabanc{\i} University under the direction of Kemal Oflazer. It is part of the BalkaNet project and the current version contains 15491 words and 14796 synonym sets \cite{Oflazer2004}. 

At this stage of the work, we were concerned only with synonymy relationship, the one that really characterizes WordNet. WordNet consists of sets of synonymous words called synsets. The words inside a synset are synonyms of each other in a symmetric manner, i.e. if word $X$ is synonym of word $Y$, then $Y$ is synonym of $X$ as well. A concept (word meaning) can be represented by the words inside its respective synonym set used to express it. For example two meanings of {\it board} can be represented unambiguously by these synonym sets: \{{\it board}, {\it plank}\} and \{{\it board}, {\it comittee}\}. In addition each synonym set is accompanied by a definition of the underlying concept, an example of its usage in a sentence or expression, the domain to which it belongs, for example the domain of the concept {\it organism} is biology.  Its lexical and computational features make WordNet an efficient and widely used tool for Natural Language Processing.
\section{Graph Construction}
We made use of WordNet as a lexical resource in order to construct our complex networks in both languages. The resulting networks showed to be scale free, have a small worlds structure and follow power law \cite{Newman2003}. First of all, we assigned a unique ID to each word available in WordNet and stored them in the appropriate tables of our database. We treated each word as a vertex (node) of the network and every synonymous relationship between two words as an edge. For example, if there is a synonymous relationship between the words {\it hard} and {\it difficult}, then there is an edge from {\it hard} to {\it difficult} as well as from {\it difficult} to {\it hard}. 

\begin{figure}[htdp]
\centering
\includegraphics[width=2.5in]{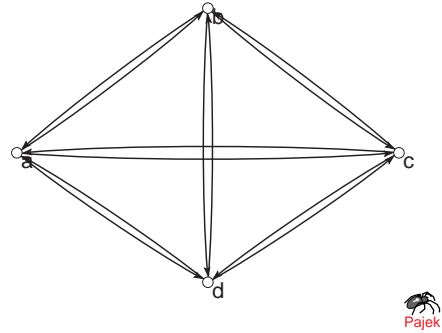}
\caption{Single Synset}
\label{figSingleSynset}
\end{figure}

Consider a synonym set $A$ consisting of the words {$a$, $b$, $c$, $d$}, then we construct twelve edges representing the symmetric synonymous relationship between every word pair as shown in Figure \ref{figSingleSynset}. The number of edges $E$, resulting from the words in a synonym set could be expressed as: 

\begin{equation}
\label{eqnEdges}
E = 2 \times {V \choose 2}
\end{equation}

where $V$ is the number of words (vertices) of the synonym set. In this way, our graphs are directed, preserving the symmetric synonymous relationship between any two words.

For every pair of synonymous words there are two cases: either corresponding meanings (named as sense in WordNet) of the words are synonymous or non-corresponding ones are. Consider this example taken from the English WordNet: the first sense (meaning) of the word {\it barely} is synonymous to the first sense of the word {\it hardly}, but the third sense of word {\it cut} is synonymous to the fourth sense of the word {\it shortened}. We don't distinguish between corresponding or non corresponding senses of synonymous words when we analyze the networks, i.e. we don't do any sense filtering during graph analysis at this stage of the work. We keep the information that a certain sense of a certain word is synonymous to a certain sense of another word in the database but we don't show this specifically in our network. Whenever there is a clique i.e. a fully connected component in the graph, resulting from nodes linked by single edges, then this means that the nodes of this component take place in the same synonym set and share the same concept (word meaning).

\begin{figure}[htdp]
\centering
\includegraphics[width=2.5in]{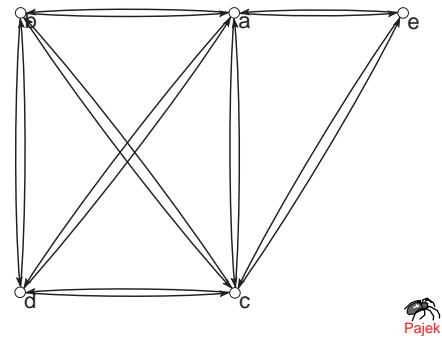}
\caption{Two Synsets}
\label{figDoubleSynsets}
\end{figure}

Often it is possible to go from one word to another one of a different meaning through a path in the network. Consider the graph in Figure \ref{figDoubleSynsets}, which results from two synonym sets. Let synonym set $A$ = \{$a$, $b$, $c$, $d$\} and set $B$ = \{$a$, $c$, $e$\}. In set $A$ the synonymous relationships with respect to the senses of the words are as follows: the first sense of $a$, the third sense of $b$, the second sense of $c$ and the fourth sense of $d$ are synonymous to each other. On the other hand, the synonymous relationships with respect to the senses of the words in set $B$ are: the third sense of $a$, the first sense of $c$ and the fifth sense of $e$ are synonymous to each other. In Figure \ref{figDoubleSynsets} set $A$ is represented by the fully connected component (clique) consisting of the nodes {$a$, $b$, $c$, $d$} and set $B$ is represented by the clique consisting of the nodes {$a$, $c$, $e$}. The concepts (word meanings) expressed by these two synonym sets are different but there exist paths from one set to the other for example: $d$-$a$-$e$, $e$-$c$-$b$ etc.
\section{Analysis Results}
%%%%%%%%%%%%%%%%%%%%%%%%%%%%%%%%%%%%%%%%%%%%%%%%%%%%%%%%%%%%%%%%%%%%%%%%%%%
% Basic Network Properties

After constructing our complex networks, we began to analyze them on basis of graph theory criteria. We made use of Pajek software for network analysis \cite{PajekSoft}. First we looked at some basic network properties, which we briefly show by Table \ref{tblBasicNetwork}:

\begin{table}
\caption{Basic Network Properties}
\centering
\begin{tabular}{|l|l|l|}
\hline
{\bf Property} & {\bf English} & {\bf Turkish} \\
\hline
The number of words &     125207 &      15491 \\
\hline
The number of edges &     271895 &      18047 \\
\hline
  Diameter & 30 (between {\it elating} & 32 (between {\it \c{c}o\v{g}almak}\\
  & and {\it well timed}) & and {\it yol katetmek})\\
\hline
Average distance among &      23924 &      17807 \\
reachable pairs && \\
\hline
\end{tabular}  
\label{tblBasicNetwork}
\end{table}

%%%%%%%%%%%%%%%%%%%%%%%%%%%%%%%%%%%%%%%%%%%%%%%%%%%%%%%%%%%%%%%%%%%%%%%%%%%
%Degree Distribution Analysis

\begin{figure}[htdp]
\centering
\includegraphics[width=3.0in]{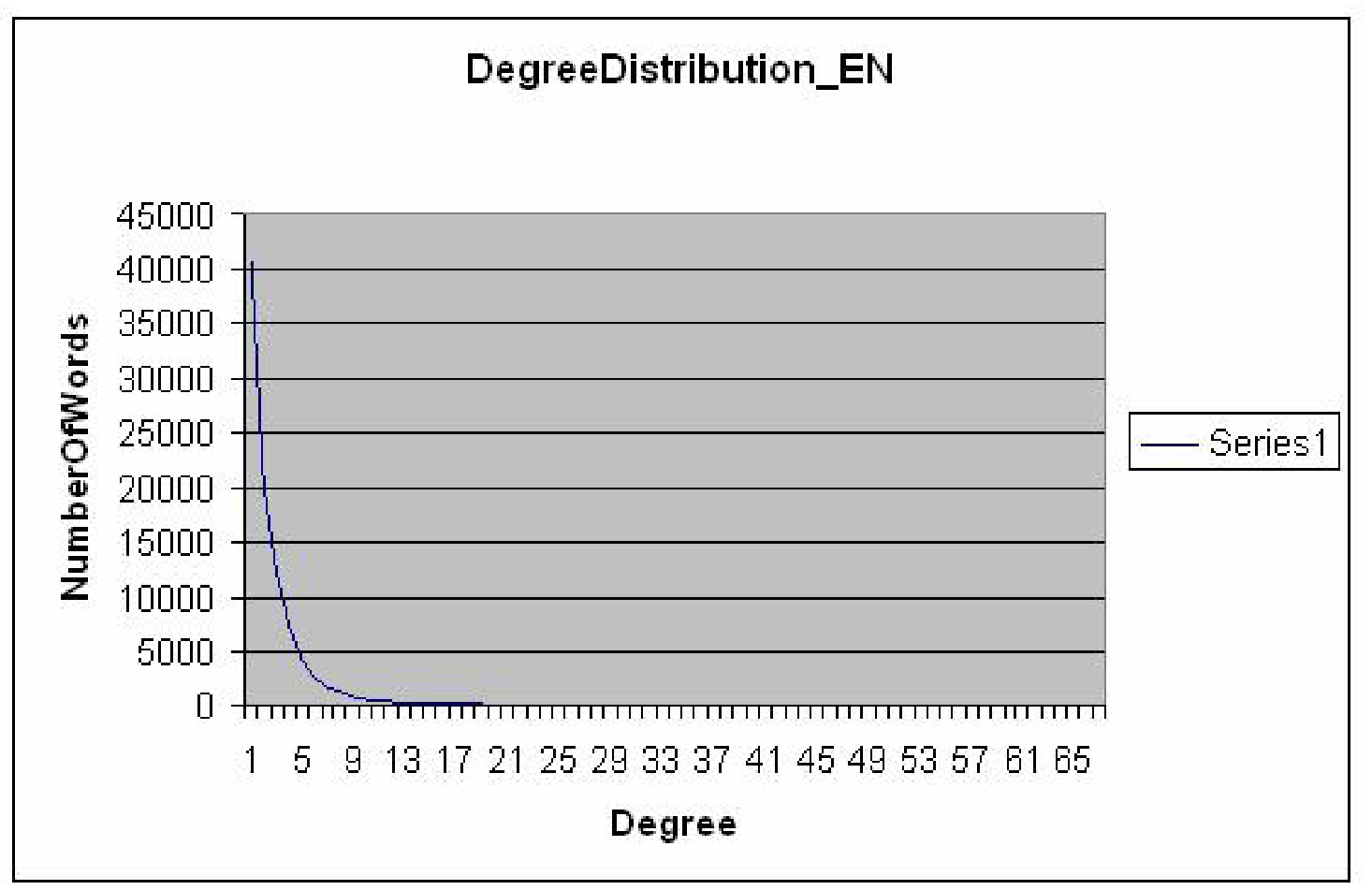}
\caption{English Degree Distribution}
\label{figEnglishDegDist}
\end{figure}

\begin{figure}[htdp]
\centering
\includegraphics[width=3.0in]{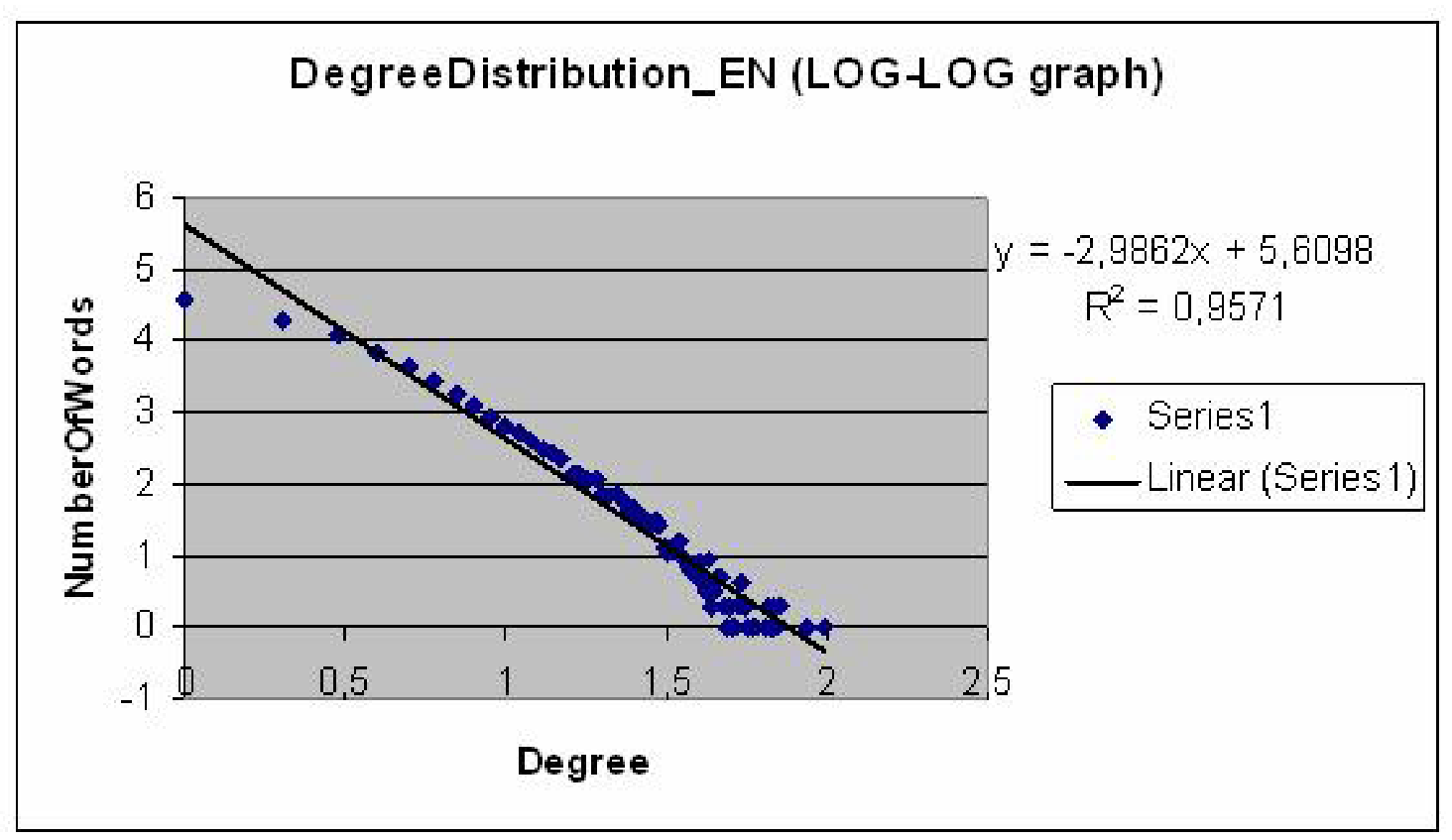}
\caption{English Degree Distribution LOG-LOG}
\label{figEnglishDegDistLOG}
\end{figure}

\begin{figure}[htdp]
\centering
\includegraphics[width=3.0in]{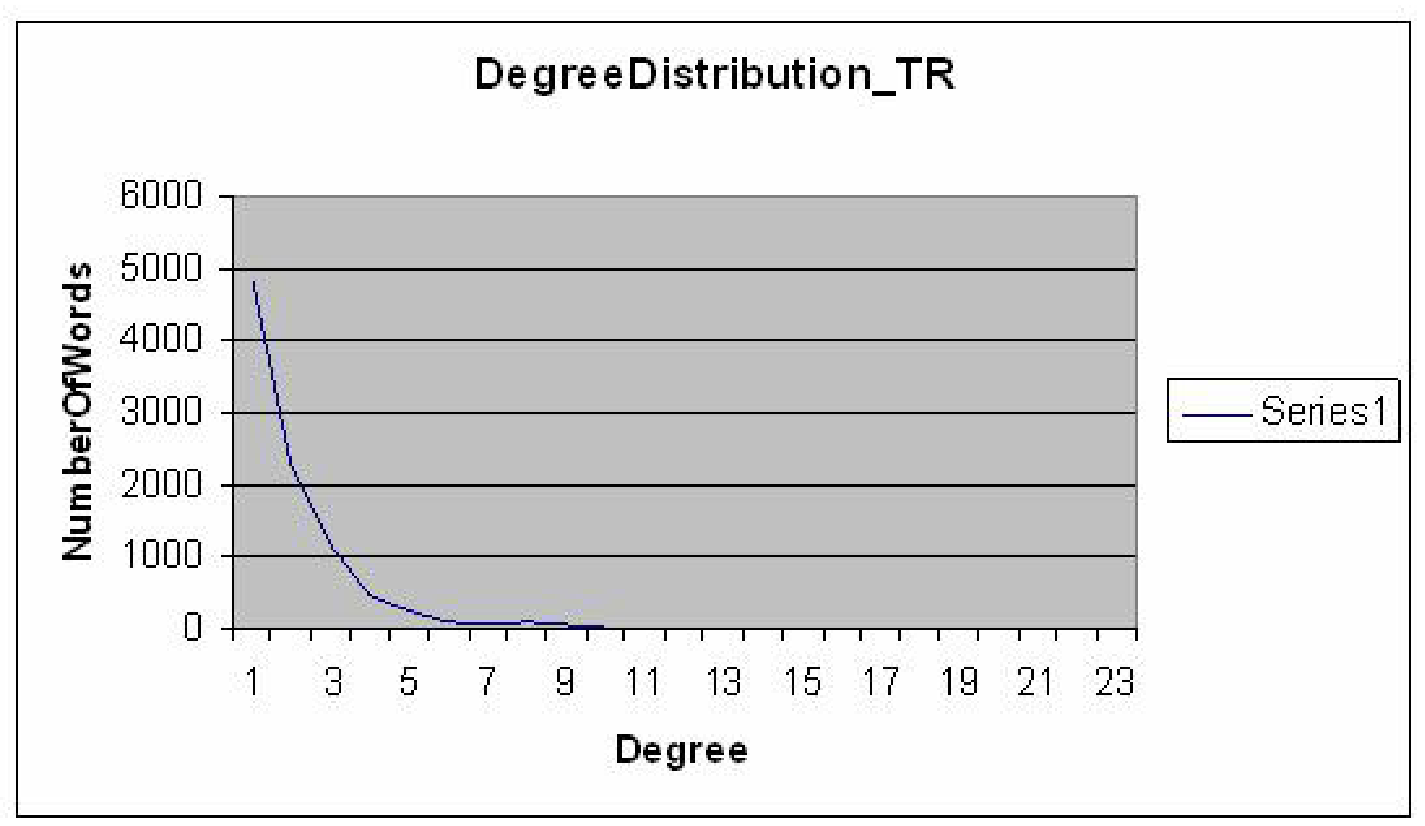}
\caption{Turkish Degree Distribution}
\label{figTurkishDegDist}
\end{figure}

\begin{figure}[htdp]
\centering
	\includegraphics[width=3.0in]{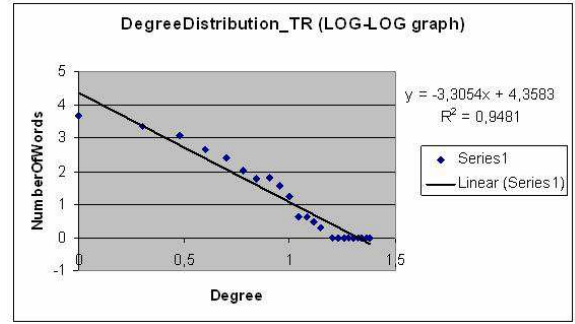}
\caption{Turkish Degree Distribution LOG-LOG}
\label{figTurkishDegDistLOG}
\end{figure}

We also examined the degree distribution in both networks and constructed the respective normal and log-log graphs. The normal graphs have a hyperbolic shape and the log-log graphs show that the data follows power law. It can be seen that the graphs for English have a clearer shape and display the data behavior better than the graphs for Turkish. This happens because the data in English WordNet is much more abundant than that in Turkish one, but still the graphs for both languages are rather similar to each other in shape and behaviour.

%%%%%%%%%%%%%%%%%%%%%%%%%%%%%%%%%%%%%%%%%%%%%%%%%%%%%%%%%%%%%%%%%%%%%%%%%%%
%TOP 20 Analysis

%TURKISH CHARACTERS
%\c{c}       \c{s}       \"{u}       \"{o}       \v{g}       {\i}
%\c{C}       \c{S}       \"{U}       \"{O}       \v{G}       

\begin{table}
\caption{TOP 20 Words with largest number of synonyms}
\centering
\begin{tabular}{|l|l|l|l|l|}
\hline
 & {\bf English} & {\bf Number of} & {\bf Turkish} & {\bf Number of } \\
 & {\bf Concept} & {\bf Synonym}   & {\bf Concept} & {\bf Synonym} \\
\hline
         1 &      break &         98 &   d\"{u}\c{s}\"{u}nmek &         24 \\
\hline
         2 &       pass &         86 &     tutmak &         20 \\
\hline
         3 &       hold &         72 &        tip &         17 \\
\hline
         4 &      check &         72 &        yer &         16 \\
\hline
         5 &        get &         70 &    istemek &         14 \\
\hline
         6 &       make &         69 &      olmak &         14 \\
\hline
         7 &       take &         69 & kar{\i}\c{s}t{\i}rmak &         13 \\
\hline
         8 &        bum &         68 &      par\c{c}a &         13 \\
\hline
         9 &        run &         67 &  g\"{o}stermek &         13 \\
\hline
        10 &       line &         66 &       kafa &         12 \\
\hline
        11 &        cut &         66 &      acaip &         12 \\
\hline
        12 &       deal &         64 &   \c{c}evirmek &         12 \\
\hline
        13 &      light &         60 &       ayak &         12 \\
\hline
        14 &        see &         59 &         i\c{s} &         11 \\
\hline
        15 &        set &         57 &    geli\c{s}me &         11 \\
\hline
        16 &      spoil &         55 &       ilgi &         11 \\
\hline
        17 &       cast &         55 &     destek &         11 \\
\hline
        18 &       beat &         54 &    hareket &         10 \\
\hline
        19 &       mark &         54 &      birey &         10 \\
\hline
        20 &         go &         54 &   yaratmak &         10 \\
\hline
\end{tabular}  
\label{tblTOP20}
\end{table}

Another criterion on which we investigated the networks was the number of synonyms of the words i.e. their degrees. Table \ref{tblTOP20} displays the first twenty words with the largest number of synonyms for both languages. As mentioned, since the beginning of the study our main interest was on the words with large synonym number as it may show that the concepts represented by those words and their synonyms are important to the culture of the nation(s) speaking the language(s) under consideration. Although when asked, the first word(s) to come to mind would generally be nouns, most of the results for English are verbs (although the noun version of most of them exists, verb version is considerably more dominant) as table \ref{tblTOP20} shows. The results for Turkish are slightly different, although the number of verbs is large, it still doesn't exceed that of nouns.

As mentioned, studies on cultural differences between East and West have been widely made since this is an interesting and challenging issue. A distinguished work on this field would be that of Richard E. Nisbett in Michigan University \cite{Nisbett_Geo,Nisbett_Masuda}. After one of his brilliant Chinese students claimed that the main difference between them was the fact that he (the student) saw the world as a circle while the professor as a line, the famous professor started a series of studies named as "the nature of thought". During this study a lot of questions raised, among which: "Why do Western children learn nouns more easily than verbs, while the contrary holds for eastern children, i.e. they learn verbs more easily than nouns?". When we observed the results of Table \ref{tblTOP20}, we noticed two things related to the above case: the word {\it line} is among the words with the largest synonym number for English and this is an interesting result considering the claim of the Chinese student which in turn motivated Nisbett for his famous work. 

Also, the dominant number of verbs with largest number of synonyms may be an answer to the above question about Western and Eastern children from a different point of view: since the verbs have a lot of synonyms, it would be more difficult for a child to learn and remember them. Of course this is only one explanation to such a wide and complex issue, there may exist more, depending on different points of view. If we observe carefully the table, we could also see some other interesting results for both languages. Consider the English words (verb version): {\it get}, {\it take}, {\it cut}, {\it deal}, {\it set}, {\it beat}, {\it mark}, and {\it hold}. They may express different characteristics or attributes of Western culture:

{\it set}: the dominant meaning of this verb is that of giving a value, attribute, state, quality, cost, etc. to something in attempt to fix and make it distinct from something else. In a way, it shows importance given to objects when providing them with a value of any kind.  

{\it mark}: the dominant meaning of this verb is that of making something distinct, generally by providing it with a feature that makes it distinguishable from the context or environment it is found. From this point of view, this verb also shows importance given to objects in attempting to make them distinct from the rest.

According to the work on culture and point of view, by Nisbett and Masuda, Westerners tend to give the main importance to objects rather than the environment, field or context while Eastern Asians are inclined to focus on the field, environment or context rather than the objects. We notice that the above explanations of the verbs {\it set} and {\it mark} support the above result by pointing out the importance given to objects, a characteristic of Westerners \cite{Nisbett_Masuda}. 

{\it beat}: this verb dominantly expresses ambition to succeed, to win, to be superior, one of the main features of capitalism which was born and is best applied in West.

{\it get}, {\it take}: both of these verbs mainly express the ambition to be in possession of, to obtain, to try to have something, again one of the features of capitalism, i.e. of West.

{\it deal}: this verb dominantly expresses the ability to agree on, to succeed in managing or arranging something. This verb has a wide use in the fields of business and politics; two areas in which West leads.

{\it hold}: this verb is used to express possession of something, again a well-known feature of West.

Now let us consider the results for Turkish: the first place of the list is occupied by the verb {\it d\"{u}\c{s}\"{u}nmek}, the English correspondent of which is {\it think}. This is an interesting result from a cognitive point of view since thinking is closely related to brain, but it also may be interpreted as a tendency of Turkish people (Easterners) towards meditation, a kind of deep thinking. A similarity with the English results would be the presence of {\it tutmak}, the English correspond of which is {\it hold} (discussed above). Another one we notice would be that of the presence of the verbs {\it make} and {\it yaratmak} (create, make, invent) among the results for both languages. Generally Westerners are considered as more innovative, creative and courageous in inventing new things, but the presence of {\it yaratmak} may show that this concept is important to Turkish people as well. The words {\it ilgi} (interest, involvement) and {\it destek} (support) may be interpreted from an emotional point of view as a tendency of Turkish people towards helping, supporting, paying attention to somebody. Other interesting words such as {\it i\c{s}} (work) and {\it geli\c{s}me} (progress) may be considered to express will for work and progress of Turkish people, while such features are usually attributed to Western people.

We made use of the online dictionaries in order to give objective definitions and explanations. Online Cambridge Dictionary was used for English \cite{CambridgeDict} and Zargan for Turkish \cite{ZarganDict}.

%%%%%%%%%%%%%%%%%%%%%%%%%%%%%%%%%%%%%%%%%%%%%%%%%%%%%%%%%%%%%%%%%%%%%%%%%%%
%Center Word
\begin{table}
\caption{Center Words in English (26095 words in island)}
\centering
\begin{tabular}{|l|l|}
\hline
{\bf English Concept} & {\bf Average Distance} \\
\hline
       get & 4.84 \\
\hline
      make & 4.85 \\
\hline
        go & 4.88 \\
\hline
      take & 4.91 \\
\hline
       run & 4.92 \\
\hline
\end{tabular}  
\label{tblCenterWordEN}
\end{table}

\begin{table}
\caption{Center Words in Turkish (420 words in island 1)}
\centering
\begin{tabular}{|l|l|}
\hline
{\bf Turkish Concept} & {\bf Average Distance} \\
\hline
             olmak & 8.32 \\
\hline
      ge\c{c}irmek & 8.36 \\
\hline
             uymak & 8.40 \\
\hline
       \c{c}arpmak & 8.41 \\
\hline
           serpmek & 8.48 \\
\hline
\end{tabular}  
\label{tblCenterWordTR1}
\end{table}

\begin{table}
\caption{Center Words in Turkish (152 words in island 2)}
\centering
\begin{tabular}{|l|l|}
\hline
{\bf Turkish Concept} & {\bf Average Distance} \\
\hline
      mesele & 4.72 \\
\hline
        konu & 5.02 \\
\hline
      i\c{s} & 5.12 \\
\hline
        dert & 5.12 \\
\hline
       sorun & 5.30 \\
\hline
\end{tabular}  
\label{tblCenterWordTR2}
\end{table}

\begin{table}
\caption{Center Words in Turkish (113 words in island 3)}
\centering
\begin{tabular}{|l|l|}
\hline
{\bf Turkish Concept} & {\bf Average Distance} \\
\hline
          ilgi & 4.17 \\
\hline
   e\v{g}lence & 4.36 \\
\hline
      bak{\i}m & 4.41 \\
\hline
        dikkat & 4.68 \\
\hline
          oyun & 4.69 \\
\hline
\end{tabular}  
\label{tblCenterWordTR3}
\end{table}

Another analysis we performed was that of finding the center words of both networks. To achieve this, we analyzed the largest island of the English network (the others are very small compared to it) and the three largest island of the Turkish network whose sizes are closed and comparable to each other. In Table \ref{tblCenterWordEN}, \ref{tblCenterWordTR1}, \ref{tblCenterWordTR2}, and \ref{tblCenterWordTR3}, we show the first five center words for each island in both networks. The average distance is calculated by the following formula:

\begin{equation}
x_{j} =  \frac{\sum_{i=1}^{n} (i\times N(i)}{m}
\end{equation}

where $x_{j}$ is the average distance for a node $j$ with respect to all of its neighbours, $i$ is the the number of hops, $n$ is the diameter of island, $N(i)$ is the number of nodes at $i${\it th} hop, and $m$ is the total number of nodes in the island.

As it can be seen from the tables, all of the English center words take place in Table \ref{tblTOP20} and this is important for us. Center words are important to a network and the fact that they are among the ones with largest synonym number emphasizes the importance of these words from our point of view also. Now consider the Turkish center words. In the first island (Table \ref{tblCenterWordTR1}), we see that center words are verbs. The verb {\it olmak}  which is the most central word of the island appears in Table \ref{tblTOP20} as well.

In the second island (Table \ref{tblCenterWordTR2}), only nouns are present. The word {\it i\c{s}} is also present in Table \ref{tblTOP20}. In the third island (Table \ref{tblCenterWordTR3}), we can see that the word {\it ilgi} is also present in Table \ref{tblTOP20}.

%%%%%%%%%%%%%%%%%%%%%%%%%%%%%%%%%%%%%%%%%%%%%%%%%%%%%%%%%%%%%%%%%%%%%%%%%%%
%Graph Reduction
\begin{figure}[htdp]
\centering
\includegraphics[width=3.0in]{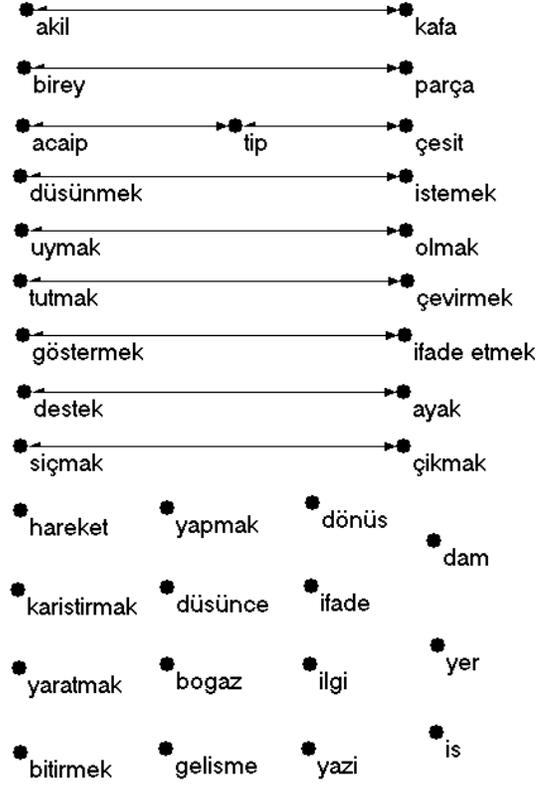}
\caption{Turkish Reduction Graph}
\label{figTurkishReduction}
\end{figure}

\begin{figure}[htdp]
\centering
\includegraphics[width=3.0in]{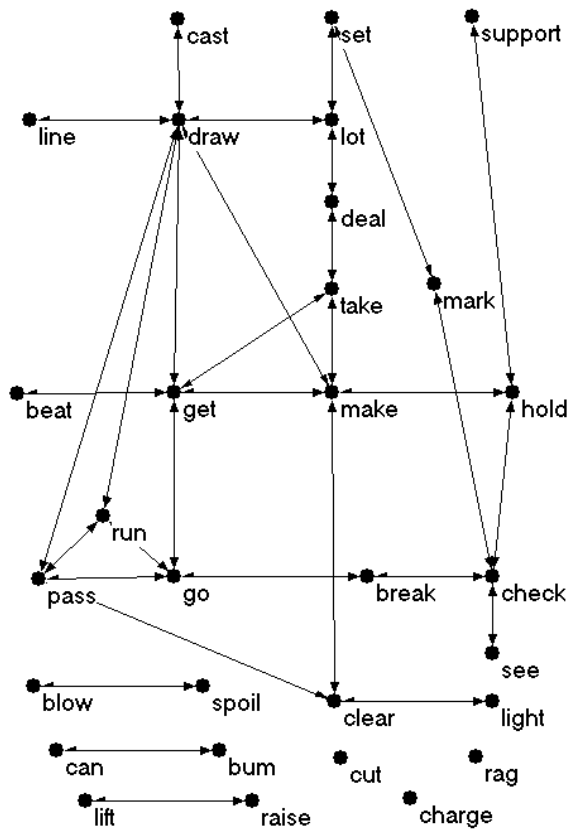}
\caption{English Reduction Graph}
\label{figEnglishReduction}
\end{figure}

In Figure \ref{figEnglishReduction} and Figure \ref{figTurkishReduction}, we give the results of another network analysis, graph reduction. We used Pajek to perform this analysis based on the following criteria: for English the words which have more than 48 synonyms and for Turkish the ones that have more than 10 synonyms should survive the graph reduction. The survivor words are the same with those in Table \ref{tblTOP20} as it is obviously expected. Also, we can notice that the reduced graph for English preserves in great part the connectivity, while the one for Turkish doesn't. This happens because of the great difference in network sizes, i.e. English network is huge compared to the Turkish one.

\section{Conclusions}
In this work we compared the synonym networks of English, a Western language, and Turkish, an Eastern one in an attempt to find possible cultural differences or similarities between the two extremes, East and West. We made use of a well-known lexical database, WordNet to construct the complex networks for both languages. As expected, these networks are free-scale, obey to Power Law and show small world effects. Previous work on linguistic networks in several aspects has been made, but we took the challenge of comparing Eastern and Western cultures by analyzing the synonym networks of two representative languages. Synonyms are an important part of a language and the need to invent new one(s) for a certain concept may differ from one culture to another, according to the importance given to that concept by those cultures. We simulated the networks on different graph theory criteria such as: degree distribution, finding center words, words with the largest degree (synonym number) and graph reduction. We obtained interesting results for both languages, especially with some English verbs and interesting words in Turkish (see Analysis Results). Most of these interesting words were found to be center words a well, emphasizing in this way their importance from both the network and our point of view. Cultural differences is a well-known and wide topic on which several work from psychologists and sociologists has been done; we approached this matter by a different and new point of view. As future work, we plan to extend our study to four other languages: Italian, Hindi, Arabic and Hebrew, the WordNet license of which we have recently obtained. In this way, the comparison would be done from a wider and more consistent perspective. We also plan to add meaning filtering to the network analysis and to include also the various relationships between synonym sets in WordNets , such as hyponymy, meronomy, antonomy, etc. in the networks.

Special thanks go to Kemal Oflazer for providing us with the Turkish WordNet.
This work was partially supported by Bogazici Research Grant BAP {\bf 07A105}.

\appendices
\vfill
\bibliographystyle{unsrt}

\begin{thebibliography}{}

\end{thebibliography}


\begin{thebibliography}{99}

\bibitem{Wordnet:Introduction}
George A.~Miller, Richard~Beckwith, Christiane Fellbaum, Derek~Gross  and Katherine Miller.
\newblock Introduction to wordnet: An on-line lexical database, August 1993.


\bibitem{Ferrer_Sole}
R.~Ferrer~i Cancho and R.~V. Sole.
\newblock The small world of human language.
\newblock {\em Proceedings of the Royal Society of London. Series B, Biological
  Sciences}, 268.1482:2261--2265, 2001.


%Buy Article Link
%http://prola.aps.org/abstract/PRE/v65/i6/e065102
\bibitem{Motter}
Motter, A.~E., de~Moura, A.~P.~S., Lai, Y.--C., \& Dasgupta, P. 
\newblock Topology of the conceptual network of language. 
\newblock {\em Physical Review E}, 65.6:065102, 2002.


\bibitem{Sigman_Cecchi}
Mariano Sigman, and Guillermo A.~Cecchi.
\newblock Global organization of the Wordnet lexicon.
\newblock {\em Proceedings of the National Academy of Sciences of the United States of America}, 99:1742--1747, 2002.

%Buy Article Link
%http://www.leaonline.com/doi/abs/10.1207/s15516709cog2901_3?cookieSet=1&journalCode=cog
\bibitem{Steyvers_Tenenbaum}
Mark Steyvers, Joshua~B. Tenenbaum.
\newblock The Large-Scale Structure of Semantic Networks: Statistical Analyses and a Model of Semantic Growth.
\newblock {\em Cognitive Science: A Multidisciplinary Journal}, 29.1:41--78, 2005.

%Google Scholar
%Online Electronic Book
%http://books.google.com/books?hl=en&lr=&id=Rehu8OOzMIMC&oi=fnd&pg=PR11&dq=fellbaum+WordNet:+An+electronic+lexical+database&ots=Ino8LlWRh4&sig=J1ze2ppOlbhQdiP5fEaA9BrRgb0#PPR15,M1
\bibitem{Fellbaum}
C.~Fellbaum.
\newblock WordNet: An electronic lexical database.
\newblock {\em Cambridge, MIT Press}, 1998.

\bibitem{Oflazer2004}
Orhan Bilgin, Ozlem Cetinoglu and Kemal Oflazer.
\newblock Morphosemantic Relations in and across Wordnets: A Study Based on Turkish.
\newblock Proceedings of the Global WordNet Conference, 2004.

\bibitem{PajekSoft}
Pajek Software.
\newblock http://vlado.fmf.uni-lj.si/pub/networks/pajek/.



\bibitem{CambridgeDict}
Cambridge Dictionaries Online.
\newblock http://dictionary.cambridge.org/.

\bibitem{ZarganDict}
Zargan Online Turkish--English Dictionary.
\newblock http://www.zargan.com/.

\bibitem{Nisbett_Masuda}
Richard E.~Nisbett, and Takahiko Masuda.
\newblock Inaugural Articles: Culture and point of view.
\newblock {\em Proceedings of the National Academy of Sciences of the United States of America}, 100:11163--11170, 2003.

%Google Scholar
%http://books.google.com/books?hl=en&lr=&id=525HX623L_cC&oi=fnd&pg=RA1-PR9&dq=nisbett+The+Geography+of+Thought:+How+Asians+and+Westerners++&ots=Rzw-Erp3gF&sig=xYKmKaSGiwM5mAZkRQG_QBrMuEc
\bibitem{Nisbett_Geo}
Nisbett, R.~E.
\newblock The Geography of Thought: How Asians and Westerners Think Differently, and Why.
\newblock {\em Free Press, New York}, 2003.

\bibitem{Newman2003}
Newman, M.~E.~J.
\newblock The Structure and Function of Complex Networks.
\newblock SIAM Review, 2003.


\end{thebibliography}

\end{document}